\documentclass[conference]{IEEEtran}
\IEEEoverridecommandlockouts
\usepackage{url}

\usepackage{cite}
\usepackage{amsmath,amssymb,amsfonts}
\usepackage{algorithmic}
\usepackage{graphicx}
\usepackage{textcomp}
\usepackage{xcolor}
\usepackage{cite}
\usepackage{amsmath,amssymb,amsfonts}
\usepackage{algorithmic}
\usepackage{graphicx}
\usepackage{textcomp}
 \usepackage[colorlinks]{hyperref}
 \hypersetup{citecolor=blue}
\usepackage[table]{xcolor}
\usepackage{balance}
\usepackage{xcolor}
\usepackage{float}  

\usepackage{fancyhdr}
\def\BibTeX{{\rm B\kern-.05em{\sc i\kern-.025em b}\kern-.08em
    T\kern-.1667em\lower.7ex\hbox{E}\kern-.125emX}}

\begin{document}

\title{\LARGE Fluid-Antenna-Aided Active User Detection With 1D-CNN \\Channel Reconstruction for Unsourced Random Access}

\author{Haoyu Liang, Zhentian Zhang, Hao Jiang, Jian Dang, Zaichen Zhang
	\thanks{ }
	\thanks{Haoyu Liang, Zaichen Zhang, Zhentian Zhang are with the National Mobile Communications Research Laboratory, Frontiers Science Center for Mobile Information Communication and Security, Southeast University, Nanjing, 210096, China. Zaichen Zhang are also with the Purple Mountain Laboratories, Nanjing 211111, China. (e-mails: \{lianghaoyu, zczhang\}@seu.edu.cn, zhentianzhangzzt@gmail.com).
	
	Jian Dang is with the National Mobile Communications Research Labo-ratory, Frontiers Science Center for Mobile Information Communication and Security, Southeast University, Nanjing 211189, China, also with the Key Laboratory of Intelligent Support Technology for Complex Environments, Ministry of Education, Nanjing University of Information Science and Tech-nology, Nanjing 210044, China, and also with Purple Mountain Laboratories, Nanjing 211111, China.(email: dangjian@seu.edu.cn). 
	
	Hao Jiang is with School of Artificial Intelligence, Nanjing University of Information Science and Technology, Nanjing 210044, China. (email: jianghao@nuist.edu.cn)}
		\thanks{This work of Jian Dang and Zaichen Zhang is partly supported by the Fundamental Research Funds for the Central Universities (2242022k60001), Basic Research Program of Jiangsu (No. BK20252003), the Key Laboratory of Intelligent Support Technology for Complex Environments, Ministry of Education, Nanjing University of Information Science and Technology (No. B2202402). The work of Hao Jiang is partly supported in part by the National Natural Science Foundation of China (NSFC) projects (No. 62471238).}
		}

\maketitle

\begin{abstract}
	In this paper, we investigate the application of fluid antenna systems (FAS) for active user detection (AUD) in unsourced random access (URA). A channel reconstruction method based on a one-dimensional convolutional neural network (1D-CNN) is proposed to effectively learn the nonlinear mapping from partial channel observations to the full channel vector. Furthermore, the reconstructed channel information is exploited to improve AUD performance via port selection. Simulation results demonstrate that the proposed 1D-CNN channel reconstructor significantly outperforms traditional methods under varying pilot lengths, achieving superior normalized mean squared error (NMSE) performance. Additionally, the reconstructed channel substantially reduces the AUD error rate compared with conventional approaches relying on traditional antenna configurations.
\end{abstract}


\section{Introduction}
With the rapid proliferation of Internet of Things (IoT) devices and the escalating demand for massive connectivity, unsourced random access (URA) \cite{URA1,URA1.11,URA1.12,URA1.2,URA1.3,URA1.4} has emerged as a promising transmission paradigm for future sixth-generation (6G) networks \cite{URA}. In URA, a large number of devices transmit messages without prior coordination, posing significant challenges for active user detection (AUD) \cite{FAA} and channel estimation at the receiver. Fluid antenna systems (FAS) \cite{fas-twc-21,fas3,fas_tutorial}, which exploit spatial reconfigurability to dynamically sample the wireless channel, have demonstrated considerable potential for improving communication performance across a variety of scenarios \cite{wt1,wt2,crb_AD,FAA,wt3,wt4,HH1,HH2,HH3,HH4,HH5}. Nevertheless, the use of FAS for AUD in URA remains insufficiently explored, particularly when accurate channel reconstruction is required under limited pilot resources and noisy observations.

Compared with conventional antenna architectures, FAS provides enhanced flexibility in exploiting spatial diversity and channel variation. Prior studies have reported its advantages in scenarios such as reconfigurable intelligent surface (RIS) assisted systems \cite{FASRIS}, integrated sensing and communication (ISAC) \cite{fas_zzt1,ISAC22}, and terahertz communications \cite{Thz}. It has also been established in \cite{URA1} that FAS can substantially improve URA performance in both single-user and multi-user settings. By switching among multiple antenna ports, FAS adaptively captures favorable channel realizations, thereby improving signal reception and mitigating interference \cite{zzt_fas3,zzt_fas4}. However, the effectiveness of FAS in URA depends critically on reliable channel reconstruction from partial observations, which remains difficult when pilot resources are limited and received signals are corrupted by noise \cite{BXU1,BXU2}.

Several methods have been developed for channel reconstruction in FAS. Representative model-based approaches include orthogonal matching pursuit (OMP) \cite{fas_omp} and selective minimum mean square error (SelMMSE) estimation \cite{fas_SelMMSE}. A two-stage successive Bayesian reconstruction framework was further proposed in \cite{fas_bayes}, where the spatial correlation of FAS channels is exploited to achieve promising reconstruction performance. Although these methods yield encouraging results, their performance often depends on model assumptions or carefully designed kernel selection, which may limit robustness and generalization in practical deployments. In contrast, deep learning techniques, particularly convolutional neural networks (CNNs), have shown strong capability for learning complex nonlinear mappings directly from data \cite{CNN_CE,wang2024,wang2025}, rendering them well-suited for FAS channel reconstruction. By capturing the spatial correlation and nonlinear characteristics of wireless channels, CNN-based methods can achieve more accurate and robust reconstruction compared with conventional model-based approaches.

Motivated by these considerations, this paper proposes a one-dimensional convolutional neural network (1D-CNN)-based channel reconstruction scheme to enhance AUD performance in FAS-enabled URA systems. First, a URA system model is formulated in which the receiver is equipped with FAS, and a switch matrix is designed to facilitate dynamic channel sampling. Then, a network comprising three 1D-CNN layers followed by two fully connected layers is developed to reconstruct the full channel state information from partial observations. The proposed method effectively learns the nonlinear mapping from partial channel observations to the complete channel vector, and the reconstructed channel is further applied to port selection. Simulation results verify the effectiveness of the proposed scheme in both channel reconstruction accuracy and active user detection performance.

\section{System Model}\label{sec.system}

\subsection{FAS-Enabled URA}

\begin{figure}[t!]
	\centering
	\includegraphics[width=\columnwidth]{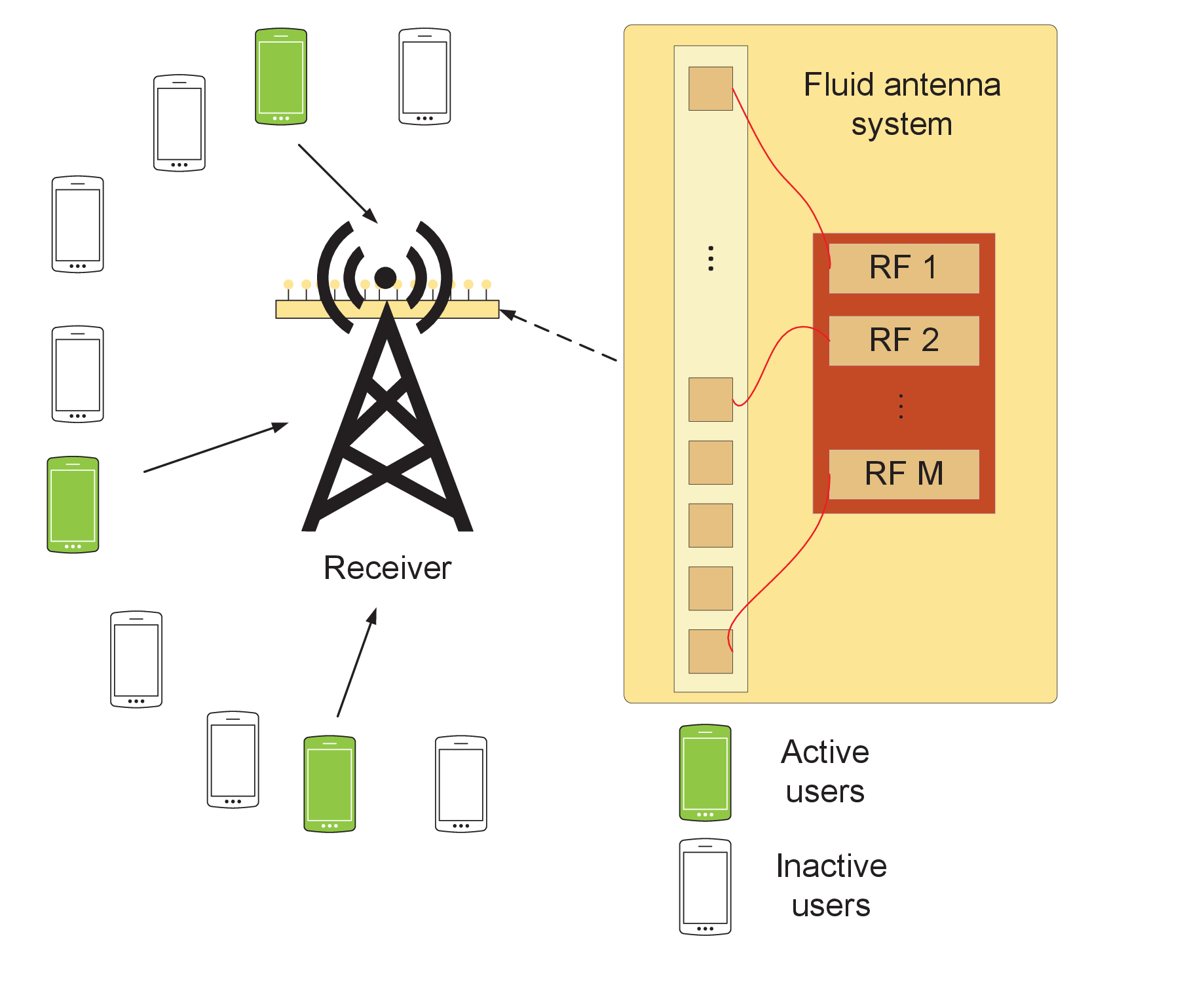}
	\caption{System model of FAS-enabled URA.}
	\label{fig.nmse}
\end{figure}

Consider a URA system with $K$ single-antenna users and a receiver equipped with an FAS. Among the $K$ users, only a small subset of $K_a \ll K$ users is active, and each active user transmits $B$ information bits over $U$ channel uses. At the receiver, $N$ ports are uniformly distributed along a fluid antenna of length $W\lambda$, where $\lambda$ denotes the carrier wavelength, as shown in Fig.~\ref{fig.nmse}. By dynamically switching among these ports, the receiver samples the spatial channel responses over the antenna aperture.

Let $L$ denote the spreading sequence length, and let $\mathbf{H}_k \in \mathbb{C}^{N \times L}$ denote the channel coefficient matrix between the $k$-th user and the receiver. The received signal is the superposition of transmissions from all active users, and is given by
\begin{equation}
	\mathbf{Y}=\sum_{k \in \mathcal{K}_a} \mathbf{H}_k \mathbf{a}_k \mathbf{x}_k\!\left(\mathbf{u}_k\right)+\mathbf{Z} \in \mathbb{C}^{N \times n},
\end{equation}
where $\mathbf{x}_k\!\left(\mathbf{u}_k\right) \in \mathbb{C}^{1 \times n}$ denotes the transmitted signal corresponding to the message bit sequence $\mathbf{u}_k \in \{0,1\}^B$, $\mathbf{a}_k \in \mathbb{C}^L$ denotes the spreading codeword selected by the $k$-th user from codebook $\mathbf{A}$, and $\mathbf{Z} \in \mathbb{C}^{N \times n}$ is the noise matrix whose entries are i.i.d. $\mathcal{CN}(0,\sigma_Z^2)$, with $\sigma_Z^2$ denoting the noise power.

The spatial reconfigurability of the FAS enables channel acquisition through dynamic port switching, which is conceptually analogous to a spatial sampling process. Assuming all $N$ ports are accessible, the receiver employs $M$ RF chains to support simultaneous port switching with negligible switching delay. A larger $M$ allows more spatial channel observations within one chip duration, at the cost of increased hardware complexity and energy consumption. Since the $M$ selected ports are mutually non-overlapping, the port switching matrix at the $l$-th chip is defined as
\begin{equation}
	\begin{aligned}
		\mathcal{S} :=
		\Big\{ \mathbf{S}_l \in \{0,1\}^{N \times M} \mid
		\sum_{n=1}^N [\mathbf{S}_l]_{n,m} = 1, \\
		\sum_{m=1}^M [\mathbf{S}_l]_{n,m} \le 1,\;
		\forall\, l \in [L] \Big\}.
	\end{aligned}
\end{equation}

At the $l$-th chip, the received signal $\mathbf{y}_l$ is expressed as
\begin{equation}
	\mathbf{y}_l = \sum_{k \in \mathcal{K}_a}
	\mathbf{S}_l^{\mathrm{H}} \mathbf{h}_{k,l}\, a_{k,l}\,
	\mathbf{x}_k\!\left(\mathbf{u}_k\right) + \mathbf{Z}.
\end{equation}

Aggregating over all $L$ chips within one coherence frame, the received signal can be written in compact form as
\begin{equation}
	\mathbf{y} = \sum_{k \in \mathcal{K}_a}
	\mathbf{S}^{\mathrm{H}} \mathbf{H}_k \mathbf{a}_k
	\mathbf{x}_k\!\left(\mathbf{u}_k\right) + \mathbf{Z},
\end{equation}
where $\mathbf{y} := \bigl[\mathbf{y}_1^{\mathrm{H}}, \ldots, \mathbf{y}_L^{\mathrm{H}}\bigr]^{\mathrm{H}}$,
$\mathbf{S} := \bigl[\mathbf{S}_1, \ldots, \mathbf{S}_L\bigr]$,
and $\mathbf{H}_k := \bigl[\mathbf{H}_{k,1}, \ldots, \mathbf{H}_{k,L}\bigr]$.

\subsection{Channel Model}

The receiver observes signals over $N$ spatially correlated ports. Let $\mathbf{h}=\bigl[h_1,h_2,\ldots,h_N\bigr]^T \in \mathbb{C}^N$ denote the channel coefficient vector across all $N$ ports. The spatial correlation is characterized by the covariance matrix
\begin{equation}
	\boldsymbol{\Sigma}=\mathbb{E}\!\left\{\mathbf{h}\mathbf{h}^H\right\} \in \mathbb{C}^{N \times N}.
\end{equation}

The channel is assumed to be fully correlated, and $\boldsymbol{\Sigma}$ is modeled as a Toeplitz matrix generated by a sinc correlation function. Specifically, the $(i,j)$-th entry of $\boldsymbol{\Sigma}$ is given by
\begin{equation}
	[\boldsymbol{\Sigma}]_{i,j}
	= \operatorname{sinc}\!\left(\frac{2\pi W}{N-1}(i-j)\right),
	\quad i,j = 1,\ldots,N,
\end{equation}
where $\operatorname{sinc}(x)=\sin(x)/x$. This Toeplitz structure captures the spatial correlation induced by the limited aperture of the FAS.

\section{Channel Reconstruction}\label{sec.proposed}

\subsection{Problem Formulation}

The objective in URA is to recover the transmitted messages $\mathbf{u}_k$, $\forall k=1,2,\ldots,K_a$, from the received signal $\mathbf{Y}$, while minimizing the required energy per bit, achieving a low AUD error rate, and meeting a target per-user probability of error (PUPE). The AUD error rate is defined as the average probability that an active user is missed, and the PUPE is defined as the average probability that an active user's message is absent from the decoded list $\hat{\mathcal{L}}(\mathbf{Y})$:
\begin{equation}
	P_e^{\mathrm{AUD}}=\mathbb{E}\left[\frac{\left|\mathcal{A}_S \backslash \hat{\mathcal{A}}_S\right|}{K_a}\right],
\end{equation}
\begin{equation}
	\mathrm{PUPE}=\frac{1}{K_a} \sum_{k \in \mathcal{K}_a} \operatorname{Pr}\left\{\mathbf{u}_k \notin \hat{\mathcal{L}}(\mathbf{Y})\right\},
\end{equation}
where $\mathcal{A}_S$ and $\hat{\mathcal{A}}_S$ denote the true and estimated active user sets, respectively. The energy per bit is expressed as
\begin{equation}
	\frac{E_b}{N_0}=\frac{nP}{c\sigma_z^2 B},
\end{equation}
where $P$ denotes the average transmit power per user per channel use, $c=2$ for real-valued channels, and $c=1$ for complex-valued channels.

Reliable CSI across all ports is essential for achieving optimal detection performance. Let $\mathbf{y} \in \mathbb{C}^{M \times 1}$ denote the observed signal and $\mathbf{h} \in \mathbb{C}^{N \times 1}$ denote the true channel vector. The channel reconstruction problem is formulated as
\begin{equation}
	\hat{\mathbf{h}}=\underset{f(\cdot),\,\mathbf{S} \in \mathcal{S}}{\operatorname{argmin}}\;\mathbb{E}\left(\|\mathbf{h}-f(\mathbf{y})\|^2\right),
\end{equation}
where $f(\cdot)$ denotes the reconstruction function, $\mathbf{S}$ denotes the port selection matrix, and $\mathcal{S}$ denotes the feasible set of selection matrices. The goal is to jointly optimize $f(\cdot)$ and $\mathbf{S}$ so as to minimize the mean squared error (MSE) between the reconstructed and true channels.

\subsection{1D-CNN Channel Reconstructor}

A shallow one-dimensional convolutional neural network (1D-CNN) is employed to learn the nonlinear mapping from partial channel observations to the full channel vector. Let $\mathbf{r} \in \mathbb{R}^{D_{\text{in}}}$ denote the real-valued input formed from the observed channel samples, and let $\hat{\mathbf{g}} \in \mathbb{R}^{D_{\text{out}}}$ denote the reconstructed channel vector. Since one-dimensional convolution requires an explicit channel dimension, the input is first reshaped as
\begin{equation}
	\mathbf{R}^{(0)} \in \mathbb{R}^{1 \times D_{\text{in}}}.
\end{equation}

The network comprises three consecutive 1D convolutional layers, each followed by batch normalization and a LeakyReLU activation. The output of the $l$-th convolutional block is given by
\begin{equation}
	\mathbf{R}^{(l)}=\phi\!\left(\mathrm{BN}^{(l)}\!\left(\operatorname{Conv}^{(l)}\!\left(\mathbf{R}^{(l-1)}\right)\right)\right), \quad l=1,2,3,
\end{equation}
where $\operatorname{Conv}^{(l)}(\cdot)$ denotes the 1D convolution in the $l$-th layer, $\mathrm{BN}^{(l)}(\cdot)$ denotes batch normalization, and $\phi(\cdot)$ is the LeakyReLU activation function, defined as
\begin{equation}
	\phi(x) = \begin{cases} x, & x \geq 0, \\ \alpha x, & x < 0, \end{cases}
\end{equation}
where $\alpha$ is a small positive constant, typically set to $0.01$. The $c$-th output feature at spatial position $i$ in the $l$-th layer is computed as
\begin{equation}
	\left[\operatorname{Conv}^{(l)}\!\left(\mathbf{R}^{(l-1)}\right)\right]_{c,i}
	= b_c^{(l)} + \sum_{c'=1}^{C_{l-1}}\sum_{m=0}^{K_l-1} w_{c,c',m}^{(l)}\, R_{c',\,i+m-p_l}^{(l-1)},
\end{equation}
where $C_{l-1}$ denotes the number of input channels in the previous layer, $K_l=3$ is the kernel size, $p_l=1$ is the zero-padding length, and $w_{c,c',m}^{(l)}$ and $b_c^{(l)}$ are the convolution kernel weights and bias, respectively.

The three convolutional layers progressively expand the channel dimension as $1 \to 16 \to 32 \to 64$, yielding
\begin{equation}
	\mathbf{R}^{(1)} \in \mathbb{R}^{16 \times D_{\text{in}}}, \quad
	\mathbf{R}^{(2)} \in \mathbb{R}^{32 \times D_{\text{in}}}, \quad
	\mathbf{R}^{(3)} \in \mathbb{R}^{64 \times D_{\text{in}}}.
\end{equation}

To compress the spatial dimension and extract a compact global representation, adaptive average pooling (AAP) is applied to the output of the third convolutional layer, yielding
\begin{equation}
	\mathbf{R}^{(4)}=\operatorname{AAP}\!\left(\mathbf{R}^{(3)}\right) \in \mathbb{R}^{64 \times 16}.
\end{equation}
The pooled feature map is then vectorized as
\begin{equation}
	\mathbf{f}=\operatorname{vec}\!\left(\mathbf{R}^{(4)}\right) \in \mathbb{R}^{1024}.
\end{equation}

This feature vector is fed into a regression head composed of two fully connected layers. The first layer projects $\mathbf{f}$ into a 256-dimensional latent space followed by a LeakyReLU activation:
\begin{equation}
	\mathbf{h}=\phi\!\left(\mathbf{W}_1 \mathbf{f}+\mathbf{b}_1\right), \quad \mathbf{h} \in \mathbb{R}^{256},
\end{equation}
where $\mathbf{W}_1$ and $\mathbf{b}_1$ denote the weight matrix and bias vector of the first fully connected layer. To improve generalization and mitigate overfitting, a dropout layer is applied to the hidden representation:
\begin{equation}
	\tilde{\mathbf{h}}=\operatorname{Dropout}(\mathbf{h}).
\end{equation}
The reconstructed channel vector is finally obtained through a second linear transformation:
\begin{equation}
	\hat{\mathbf{g}}=\mathbf{W}_2\tilde{\mathbf{h}}+\mathbf{b}_2, \quad \hat{\mathbf{g}} \in \mathbb{R}^{D_{\text{out}}},
\end{equation}
where $\mathbf{W}_2$ and $\mathbf{b}_2$ denote the parameters of the output layer. The regression head thus maps the extracted high-level features to the target output space, producing the final reconstructed channel vector.

The overall reconstruction process can be compactly expressed as
\begin{equation}
	\hat{\mathbf{g}}=f_{\Theta}(\mathbf{r}),
\end{equation}
where $f_{\Theta}(\cdot)$ denotes the nonlinear mapping implemented by the 1D-CNN, and $\Theta$ denotes the full set of trainable parameters.

\begin{figure}[t!]  
\centering
\includegraphics[width=\columnwidth]{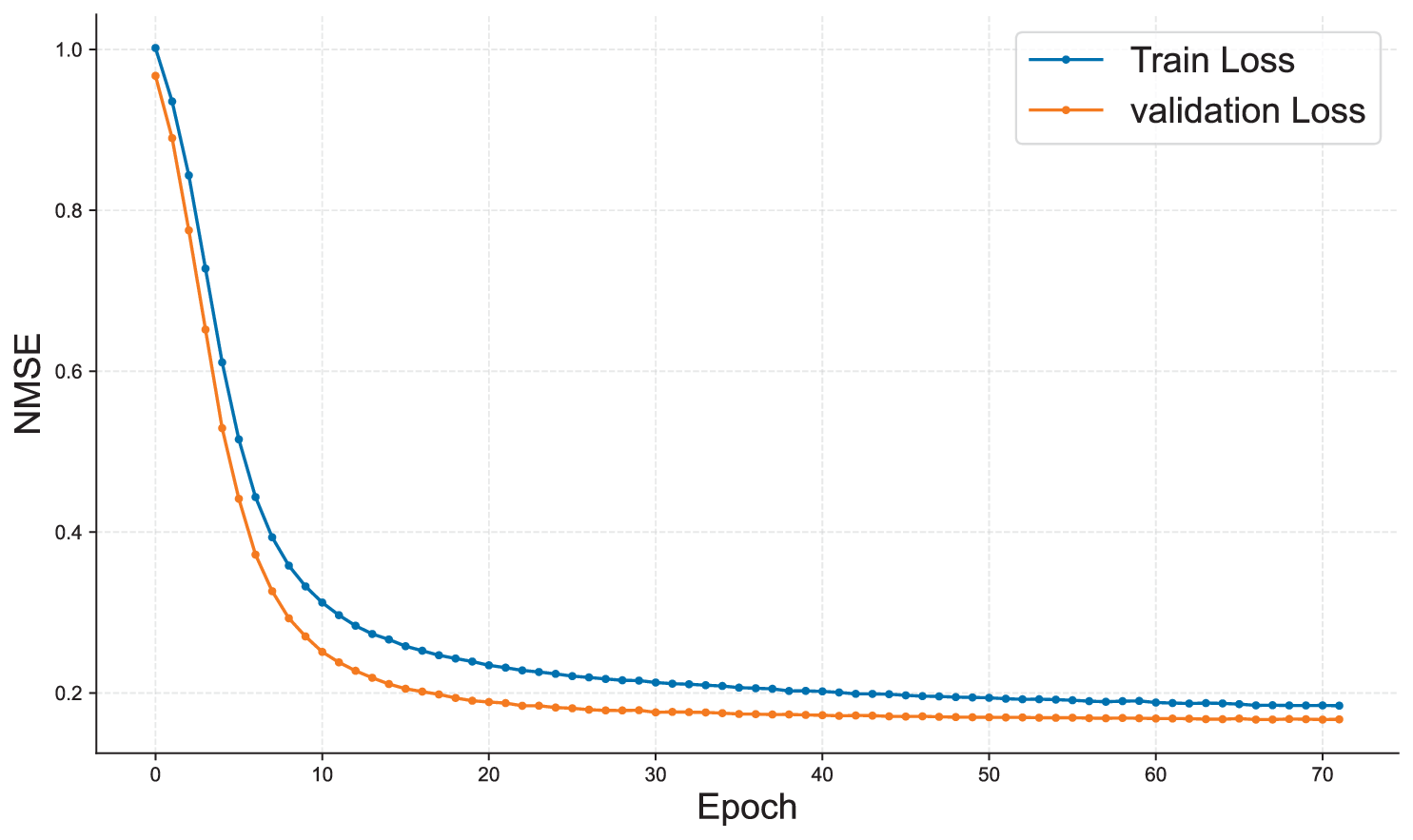}
\caption{Training and Validation NMSE for the Proposed 1D CNN-Based Channel Reconstruction Model}
\label{fig2.nmse}
\end{figure}

\begin{figure}[t!]  
\centering
\includegraphics[width=\columnwidth]{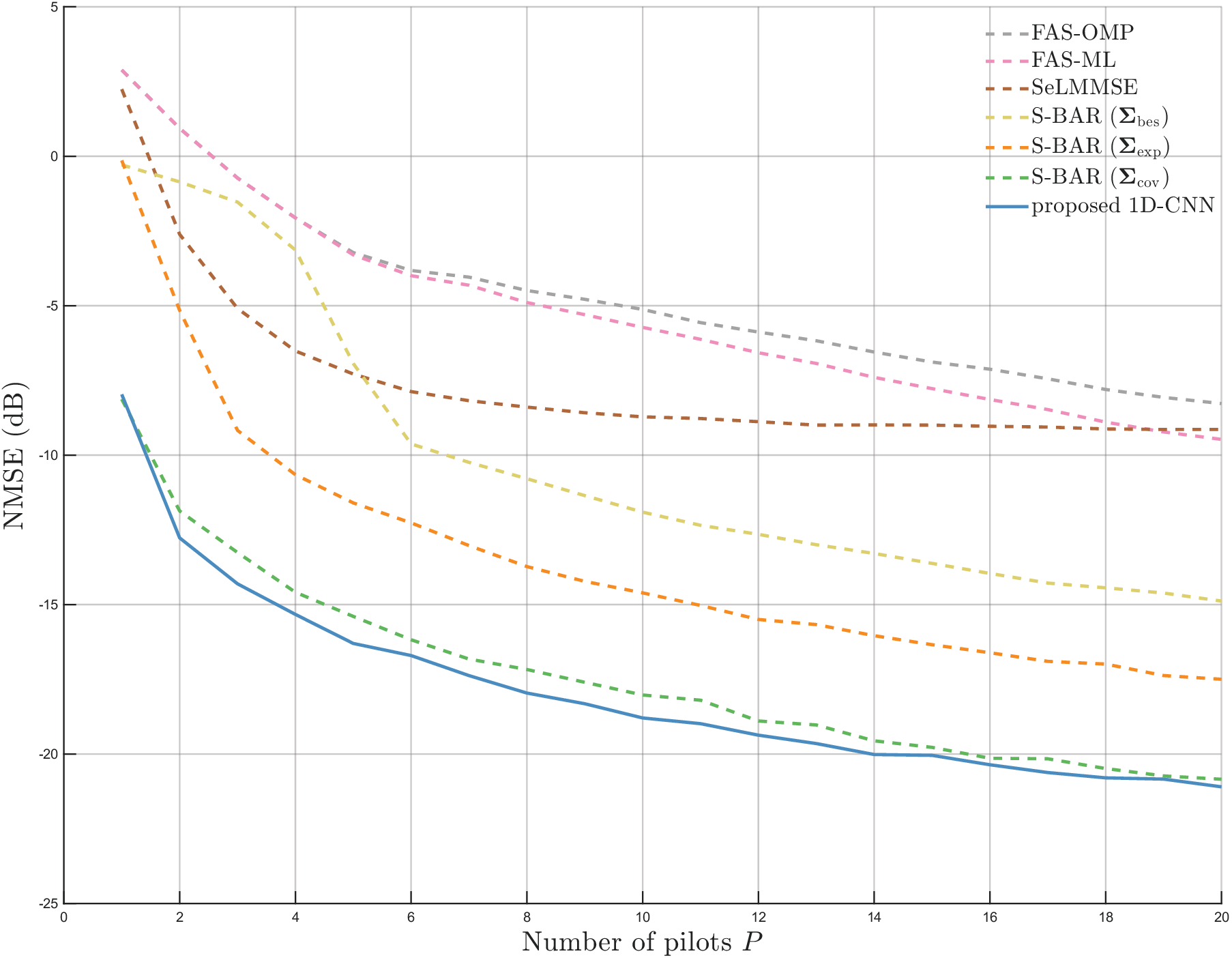}
\caption{NMSE Comparison of Different Channel Reconstruction Methods under Varying Pilot Numbers}
\label{fig3.nmse}
\end{figure}

\section{Simulation Results}
We first evaluate the proposed 1D-CNN channel reconstructor, then assess its utility for AUD against conventional antenna-based methods. The training and validation sets contain 100,000 and 5,000 samples, respectively. The input dimension is $D_{in}=2M$, where $M$ denotes the number of selected ports, and the output dimension is $D_{out}=2N$ across all $N$ ports. The pilot sequence length is $B_p=14$, with $K_a=50$ active users and $K=2^{B_p}$ total users.

\subsection{Performance of the 1D-CNN Channel Reconstructor}\label{sec.numerical}

\begin{figure}[t!]
	\centering
	\includegraphics[width=\columnwidth]{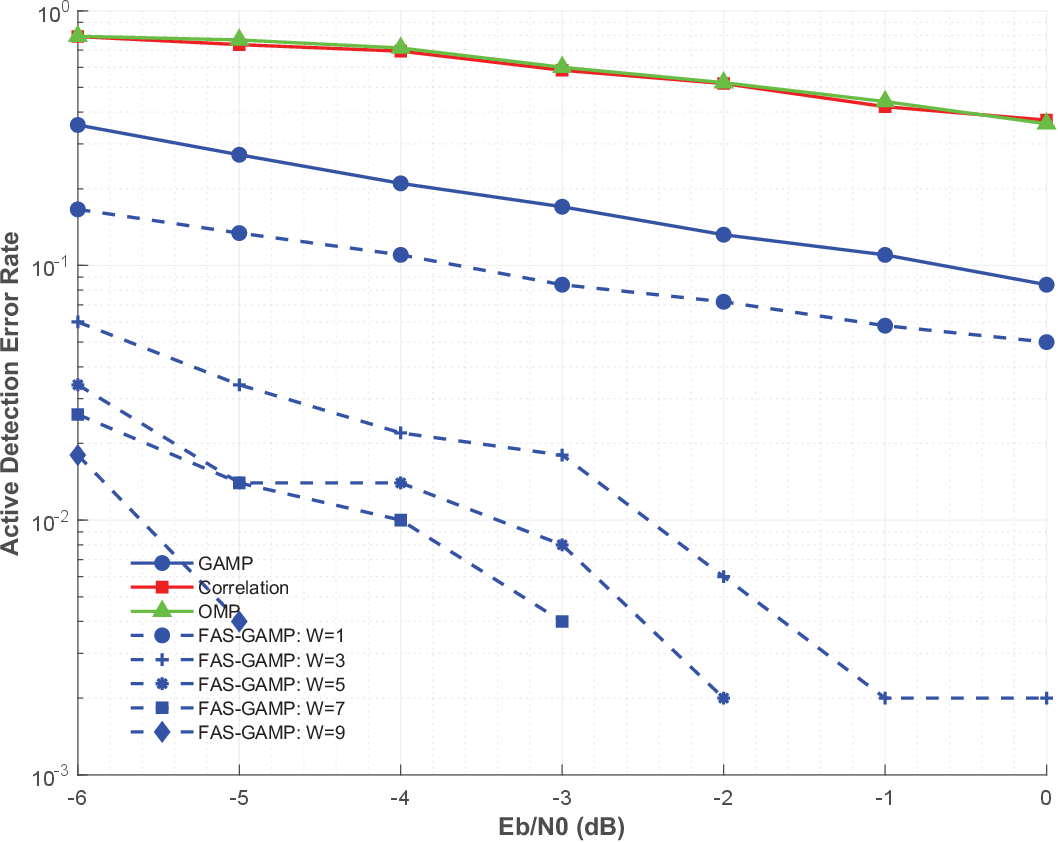}
	\caption{Performance comparison of active user detection: traditional antenna vs. FAS-GAMP with different $W$ under various SNRs}
	\label{fig4.nmse}
\end{figure}

\begin{figure}[t!]
	\centering
	\includegraphics[width=\columnwidth]{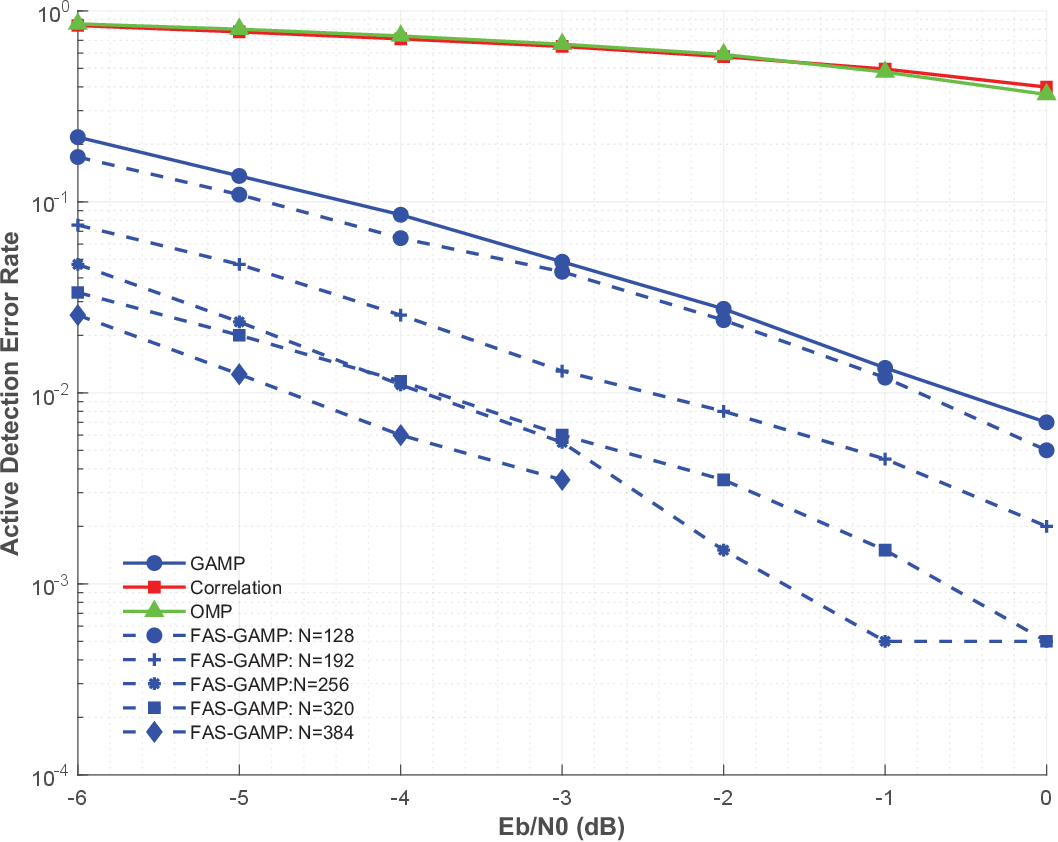}
	\caption{Performance comparison of active user detection: traditional antenna vs. FAS-GAMP with different $N$ under various SNRs}
	\label{fig5.nmse}
\end{figure}

As shown in Fig.~\ref{fig2.nmse}, both training and validation NMSE curves exhibit a consistent downward trend, confirming that the 1D-CNN effectively learns channel reconstruction from partial observations. The training NMSE decreases rapidly in the early stage (before approximately the eighth epoch) and subsequently stabilizes, while the validation curve closely tracks the training curve, indicating good generalization without noticeable overfitting. The final validation NMSE demonstrates accurate reconstruction even under severe noise conditions.

We further compare the proposed method against FAS-OMP \cite{fas_omp}, FAS-ML \cite{fas_bayes}, SelMMSE \cite{fas_SelMMSE}, and S-BAR \cite{fas_bayes}. As shown in Fig.~\ref{fig3.nmse}, the 1D-CNN consistently outperforms FAS-OMP, FAS-ML, and SelMMSE across all pilot settings, and also surpasses the best-kernel variant of S-BAR. These results confirm the advantage of deep learning for FAS channel reconstruction, particularly under limited pilot resources.

\subsection{Performance of Port Selection for AUD}

The reconstructed channel guides port selection in URA by directing the receiver to the port with the highest gain. As shown in Figs.~\ref{fig4.nmse} and \ref{fig5.nmse}, FAS-GAMP substantially outperforms OMP \cite{OMP}, correlation-based detection \cite{correlation}, and GAMP with conventional fixed antennas across the entire $E_b/N_0$ range. For a fixed $N$, increasing $W$ significantly reduces the AUD error rate, as a longer antenna provides greater spatial degrees of freedom and a higher probability of selecting a high-gain port. For a fixed $W$, increasing $N$ further reduces the error rate by providing more selection opportunities and higher received signal gain. These results confirm the importance of port selection in FAS-enabled URA and demonstrate that the reconstructed channel effectively guides this process.

\section{Conclusion}\label{sec.conclusion}
In this paper, a 1D-CNN-based channel reconstruction scheme is proposed for FAS in URA scenarios. The proposed method learns the nonlinear mapping from partial observations to the full channel vector, achieving superior NMSE performance over conventional baselines under varying pilot conditions. Simulation results verify that the reconstructed channel significantly improves AUD performance, with greater gains observed for larger antenna lengths and higher port counts. Future work will investigate more advanced architectures and training strategies to further enhance reconstruction accuracy and robustness in practical FAS deployments.

\balance


\begin{thebibliography}{99}
\bibitem{URA1}
Z.~Zhang \emph{et~al.},
``On fundamental limits of slow-fluid antenna multiple access for unsourced random access,''
\emph{IEEE Wireless Commun. Lett.},
vol.~14, no.~11, pp.~3455--3459,
Nov. 2025.
\bibitem{URA1.11}
M. J. Ahmadi, \emph{et~al.},``Integrated sensing and communications for unsourced random access: Fundamental limits and practical model,'' \emph{IEEE Trans. Wireless Commun.}, vol. 25, pp. 16547-16561, 2026.
\bibitem{URA1.12}
Z.~Zhang ,\emph{et~al.}, ``Integrated sensing and communications for unsourced random access via spectrum sharing compressive sensing approach with massive MIMO receiver,'' \emph{IEEE Trans. Veh. Technol.}, Early Access, DOI: 10.1109/TVT.2026.3657304.
\bibitem{URA1.2}
Z. Zhang, \emph{et al.}, ``Unsourced random access via random scattering with turbo probabilistic data association detector and treating collision as interference,'' \emph{IEEE Trans. Wireless Commun.}, vol. 23, no. 12, pp. 17899-17914, Dec. 2024. 
\bibitem{URA1.3}
Z. Zhang, \emph{et al.}, ``Efficient ODMA for unsourced random access in MIMO and hybrid massive MIMO,'' \emph{IEEE Internet Things J.}, vol. 11, no. 23, pp. 38846-38860, 1 Dec.1, 2024. 
\bibitem{URA1.4}
Z. Zhang, \emph{et al.}, ``Joint pattern, data and channel estimation for unsourced random access in GMAC and MIMO systems,'' \emph{IEEE Trans. Wireless Commun.}, Early Access, DOI: 10.1109/TWC.2026.3705749



\bibitem{URA}

Z. Zhang, \emph{et al.}, ``Joint activity detection and channel estimation for fluid antenna system exploiting geographical and angular information," \emph{IEEE J. Sel. Top. Signal Process.}, Early Access, doi: 10.1109/JSTSP.2026.3673148.
\bibitem{fas-twc-21}
K. K. Wong, A. Shojaeifard, K.-F. Tong and Y. Zhang, ``Fluid antenna systems," \emph{IEEE Trans. Wireless Commun.}, vol. 20, no. 3, pp. 1950--1962, Mar. 2021.
\bibitem{fas3}
K. K. Wong, A. Shojaeifard, K. F. Tong, and Y. Zhang, ``Performance limits of fluid antenna systems," \emph{IEEE Commun. Lett.}, vol. 24, no. 11, pp. 2469--2472, Nov. 2020.
\bibitem{wt1}
T. Wu, {\em et al.}, ``Variable block-correlation modeling and optimization for secrecy analysis in fluid antenna systems,'' {\em IEEE Trans. Wireless Commun.}, vol. 25, pp. 15069-15085, 2026.
\bibitem{wt2}
T. Wu, {\em et al.}, ``Scalable fluid antenna systems: A new paradigm for array signal processing,'' {\em IEEE J. Sel. Topics Signal Process.}, Early Access, 2026, \url{DOI: 10.1109/JSTSP.2026.3673981}.
\bibitem{crb_AD}
Z. Zhang, \emph{et al.}, ``Finite-aperture fluid antenna array design: Analysis and algorithm,'' \emph{IEEE Wireless Commun. Lett.}, vol. 15, pp. 3199-3203, 2026.
\bibitem{FAA}
Z. Zhang, \emph{et al.}, ``Cramer-rao bounds for activity detection in conventional and fluid antenna systems'', \emph{IEEE Wireless Commun. Lett.}, vol. 15, pp. 3059-3063, 2026.
\bibitem{wt3}
T. Wu, {\em et al.}, ``Fluid antenna systems enabling 6G: Principles, applications, and research directions'', {\em IEEE Wireless Commun.}, Early Access, 2025, \url{doi: 10.1109/MWC.2025.3629597 2025}.
\bibitem{wt4}
T. Wu et al., ``Reimagining wireless connectivity: The FAS-RIS synergy for 6G smart cities,'' {\em IEEE Commun. Mag.}, DOI: 10.1109/MCOM.001.2500802.
\bibitem{HH1}
H. Hong  \emph{et al.}, ``Fluid antenna system-assisted self-interference cancellation for in-band full duplex communications,'' {\em IEEE Trans. Wireless Commun.}, vol. 25, pp. 7476-7489, 2026.
\bibitem{HH2}
H. Hong \emph{et al.}, ``FAS meets OFDM: Enabling wideband 5G NR,'' {\em IEEE Trans. Commun.}, vol. 73, no. 11, pp. 12884--12898, Nov. 2025.
\bibitem{HH3}
H. Hong \emph{et al.}, ``Downlink OFDM-FAMA in 5G-NR systems,'' {\em IEEE Trans. Wireless Commun.}, vol. 24, no. 12, pp. 10116--10132, Dec. 2025.
\bibitem{HH4}
H. Hong, K. -K. Wong, K. -F. Tong, H. Shin and Y. Zhang, ``Coded fluid antenna multiple access over fast fading channels,'' {\em IEEE Wireless Commun. Lett.}, vol. 14, no. 4, pp. 1249--1253, April 2025.
\bibitem{HH5}
H. Hong \emph{et al.}, ``Fluid antenna multiple access for 6G: A holistic review,'' {\em IEEE Open J. Commun. Soc.}, vol. 7, pp. 2607--2633, 2026.

\bibitem{fas_tutorial}
W. K. New {\em et al.}, ``A tutorial on fluid antenna system for 6G networks: Encompassing communication theory, optimization methods and hardware designs,'' \emph{IEEE Commun. Surv. Tuts.}, vol. 27, no. 4, pp. 2325--2377, Aug. 2025
\bibitem{FASRIS}
Z. Zhang, \emph{et al.}, ``Fundamental tradeoffs for multiple access in finite-blocklength regime,'' accepted to in {\em Proc. IEEE Int. Conf. Computer Communication and Artificial Intelligence (CCAI)}, preprint: arXiv: 2601.05165, 2026.

\bibitem{fas_zzt1}
Z. Zhang, \emph{et al.}, ``On fundamental limits for fluid antenna-assisted integrated sensing and communications for unsourced random access," \emph{IEEE J. Sel. Areas Commun.}, vol. 44, pp. 136-149, 2026.
\bibitem{ISAC22}
Y. Zhang \emph{et~al.}, ``Backscatter device-aided integrated sensing and communication: A pareto optimization framework,'' \emph{IEEE Trans. Wireless Commun.}, vol. 25, pp. 17958-17974, 2026.
\bibitem{Thz}
L. Tlebaldiyeva, \emph{et al.}, "Outage Performance of Fluid Antenna System (FAS)-aided Terahertz Communication Networks," in {\em Proc. IEEE Int. Conf.
Commun. (ICC)}, Rome, Italy, 2023, pp. 1922-1927.


\bibitem{zzt_fas3}
Z. Zhang, \emph{et al.}, ``Finite-blocklength fluid antenna systems with spatial block-correlation channel model," \emph{IEEE Wireless Commun. Lett.}, vol. 15, pp. 1911-1915, 2026. 

\bibitem{zzt_fas4}
Z. Zhang, {\em et al.}, ``Coded pattern unsourced random access with analyses on sparse pattern demapper,'' \emph{IEEE Trans. Wireless Commun.}, vol. 25, pp. 5306-5319, 2026.
\bibitem{BXU1}
B. Xu, \emph{et~al.}, ``Channel estimation for Rydberg atomic receivers,'' {\em IEEE Wirel. Commun. Lett.}, vol. 14, no. 9, pp. 2957–2961, Jun. 2025.
\bibitem{BXU2}
B. Xu, \emph{et~al.}, ``Resource allocation for near-field communications: Fundamentals, tools, and outlooks,'' {\em IEEE Wireless Commun.}, vol. 31, no. 5, pp. 42–50, Jul. 2024.

\bibitem{fas_omp}
H. Hong \emph{et al.}, ``A contemporary survey on fluid antenna systems: fundamentals and networking perspectives,''  {\em  IEEE Trans. Netw. Sci. Eng.},  vol. 13, pp. 2305--2328, 2026.


\bibitem{fas_SelMMSE}
C. Skouroumounis, {\em et al.}, ``Fluid antenna with linear MMSE channel estimation for large-scale cellular networks,'' {\em IEEE Trans. Commun.}, vol. 71, no. 2, pp. 1112–1125, Feb. 2023.

\bibitem{fas_bayes}
Z. Zhang, \emph{et al.}, ``Successive Bayesian Reconstructor for Channel Estimation in Fluid Antenna Systems,'' \emph{IEEE Trans. Wireless Commun.}, vol. 24, no. 3, pp. 1992-2006, March 2025.

\bibitem{CNN_CE}
B. Lin, {\em et al.}, ``A Novel OFDM Autoencoder Featuring CNN-Based Channel Estimation for Internet of Vessels,'' \emph{IEEE Internet Things J.}, vol. 7, no. 8, pp. 7601-7611, Aug. 2020.
\bibitem{wang2025}
C. Wang, {\em et al.}, ``Large Language Model Empowered Design of Fluid Antenna Systems: Challenges, Frameworks, and Case Studies for 6G,'' {\em IEEE Wireless Commun.}, Early Access, DOI: 10.1109/MWC.2025.3600949.

\bibitem{wang2024}
C. Wang, {\em et al.}, ``AI-Empowered Fluid Antenna Systems: Opportunities, Challenges, and Future Directions,'' {\em IEEE Wireless Commun.}, vol. 31, no. 5, pp. 34-41, October 2024.
\bibitem{OMP}
V. K. Amalladinne, {\em et al.}, ``A coded compressed sensing scheme for unsourced multiple access,'' {\em IEEE Trans. Inf. Theory}, vol. 66, no. 10, pp. 6509–6533, Oct. 2020.

\bibitem{correlation}
A. K. Pradhan, {\em et al.}, ``LDPC codes with soft interference cancellation for uncoordinated unsourced multiple access,'' in {\em Proc. IEEE Int. Conf.
Commun. (ICC)}, Jun. 2021, pp. 1–6.



\end{thebibliography}
\end{document}